\documentclass[aps,prb,preprintnumbers,groupedaddress,twocolumn] {revtex4-1}
\usepackage{graphicx, float}
\usepackage{dcolumn}
\usepackage{bm}
\usepackage{amsmath}
\usepackage{amsfonts}
\usepackage{amssymb}
\usepackage{graphicx}
\usepackage{dcolumn}
\usepackage{bm}
\usepackage{amsmath,amsthm}
\usepackage{amssymb}
\usepackage{graphicx,color}
\usepackage{epstopdf}
\usepackage{soul}
\usepackage{hyperref}
\usepackage{float}
\graphicspath{{:Figs:}}
\newcommand*\Eval[3]{\left.#1\right\rvert_{#2}^{#3}}

 \def\be{\begin{equation}}
 \def\ee{\end{equation}}
 \def\bes{\begin{eqnarray}}
 \def\ees{\end{eqnarray}}

 \def\2{\frac{1}{2}}
 \def\4{\frac{1}{4}}

\begin{document}
\title{Quantum state atomic force microscopy}
\author{Ali Passian} \email{passianan@ornl.gov}
\affiliation{Computational Sciences and Engineering Division, Oak Ridge National Laboratory, Oak Ridge, TN
37831-6123 USA} 
\author{George Siopsis} \email{siopsis@tennessee.edu}
\affiliation{Department of Physics and Astronomy, The University of Tennessee, Knoxville, TN 37996-1200, USA}
\date{\today}
\begin{abstract}
New classical modalities of atomic force microscopy continue to emerge to achieve higher spatial, spectral, and temporal resolution for nanometrology of materials.
 Here, we introduce the concept of a quantum mechanical modality that capitalizes on squeezed states of probe displacement. 
 We show that such squeezing is enabled nanomechanically when the probe enters the van der Waals regime of interaction with a sample. The effect is studied in the non-contact mode, where we consider the parameter domains characterizing the attractive regime of the probe-sample interaction force.   
 \end{abstract}
 \maketitle

\section{Introduction}
In the realm of advanced measurement science and technology, label free and nondestructive  nanometrology has gained considerable attention in part due to progress in nanofabrication and nanotechnology.  The notion of picotechnology has already been contemplated~\cite{pico}. Scanning probe microscopy will undoubtedly continue to play a crucial role at such length scales. In particular, atomic force microscopy (AFM)~\cite{msafm}, allowing for highly controlled and precise actuation, excitation, and manipulation of a microcantilever probe and a given sample region, is known to be minimally invasive and thus ideal for a variety of applications and material characterization~\cite{hpfm}.  Such force metrology may involve magnitudes as small as sub atto Newton in the case of AFM~\cite{rugar}, or other approaches~\cite{moser,gavartin,ranjit}.

\begin{figure}
    \centering
            \includegraphics[scale=.55]{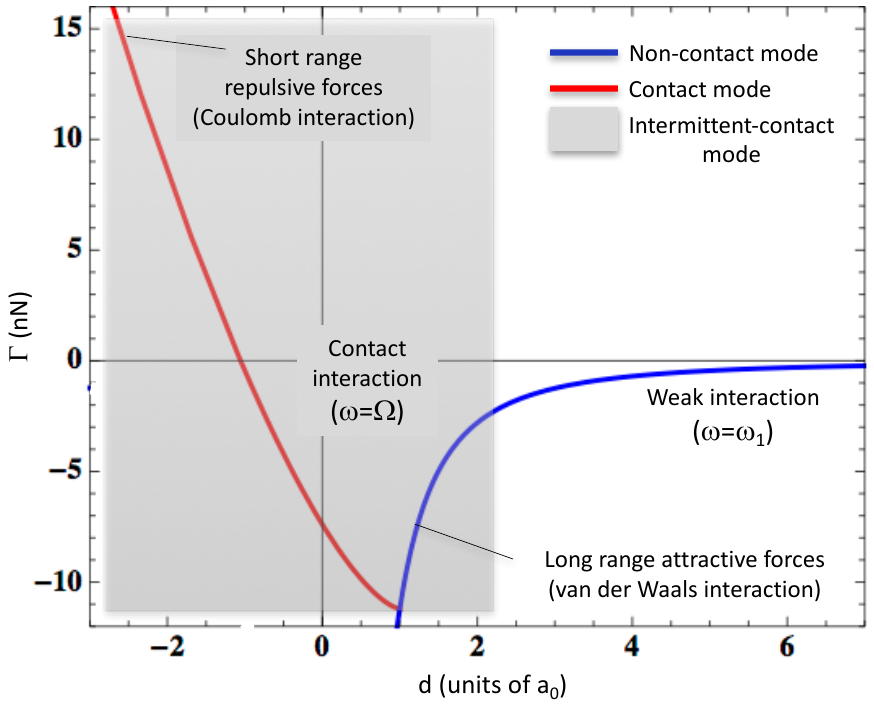}
           \caption[]{The vdw-DMT force regimes considered in the QAFM. The nonlinearity in the force is explored for obtaining quantum squeezing in the deformation state of the probe. At a distance of about the interatomic separation $\propto a_0$, that is, about the length of a chemical bond, atoms are essentially in contact. The discontinuity in the force transition region may be smoothed by employing polynomial fitting functions.   }
    \label{vdw}
\end{figure}

Here we propose the notion of a quantum enabled atomic force microscopy (QAFM), which operates on the premise of the ability of
the probe-sample interaction force to produce a squeezed quantum state for the cantilever probe. 
We propose to read out such a squeezed state by forming a probe-based cavity~\cite{aspelmeyer,walls:milburn}. Minute lateral displacements of the
sample engender  a variation in the tip-sample force and a feedback loop subsequently adjusts the separation distance to obtain a displacement set-point. Thus, by default, the readout approach in QAFM is fundamentally different than standard  techniques including the highly sensitive interferometric and phase sensitive displacement measurements. 
Through this process, for example, topographic maps with extremely high resolution are envisioned to be obtained. In this article, we aim to explore the feasibility of the van der Waals forces (Fig.~\ref{vdw}), prevailing in the nanometers region of the AFM probe tip-sample surface distance as depicted in Fig.~\ref{setup}, to engender sufficient quantum squeezing in the deformation state of the probe.        
Recent studies have demonstrated mechanical squeezing via strong optomechanical interactions~\cite{nat}. Here, we explore a different 
nonlinear interaction that results in mechanical squeezing without strong optical interactions, although optical interactions are used for the purpose
of reading out the squeezing. 
The  introduced concept of mechanical squeezing here can
potentially lead to force microscopy beyond the classical boundaries. 
Within the context of improving the measurement sensitivity and precision, optical or mechanical squeezing based on other methods have been proposed or demonstrated~\cite{nat,ligo,taylor,lecocq}. The scanning probe microscopy suite provides powerful technologies for metrology. Whereas for each sample one approach may be better than other, the AFM has proved to be highly inclusive (\emph{e.g.}, not limited to only electrically or optically conductive materials etc). Force microscopy with AFM can yield a lateral resolution of about 1~nm and a depth resolution better than about 0.1~nm.

Our discussion is organized as follows. In section \ref{sec:2}, we define the interaction force between the two macroscopic bodies, the probe and the sample, based on the van der Waals and Derjaguin, Muller, Toporov  (vdw-DMT) force. 
Employing this force, in section \ref{sec:3}, we describe the probe dynamics within the Euler-Bernuoulli model, and in section \ref{sec:q}, obtain the quantization of the calculated deformation state of the probe and the degree of squeezing. Conclusions are provided in section \ref{sec:conc}.  

\section {Probe-sample interaction model}
\label{sec:2}
\begin{figure}
    \centering
           \includegraphics[width=8.0cm]{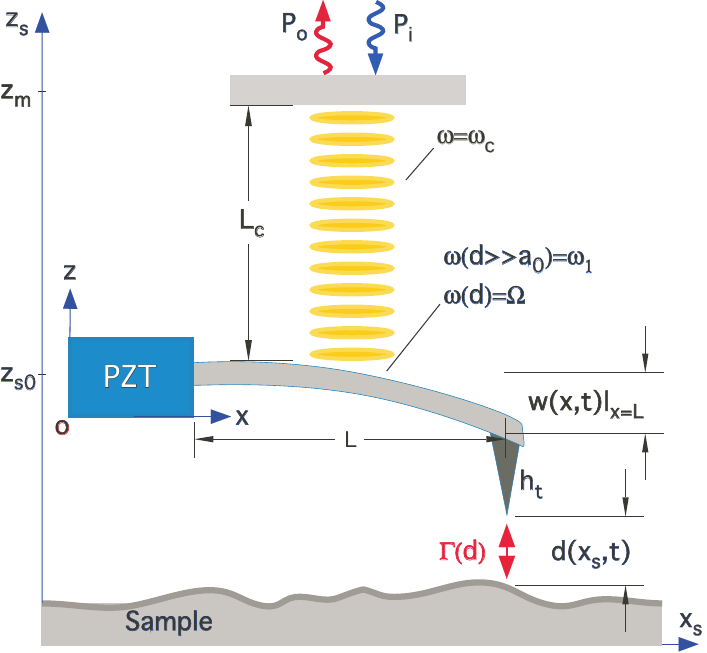}
           \caption[]{
           High resolution material characterization by use of van der Waals force induced quantum squeezed deformation state of the probe. The probe of length $L$ and a fixed mirror a distance $L_c$ above it at $z_m$, and an input beam $P_i$ forms a cavity with a frequency $\omega_c$. At a sample location $x_s$, within a distance $d$ of the attractive regime of the tip-sample interaction domain, the probe oscillating with a frequency $\Omega$, exhibits amplitude squeezing, which is detected from the outgoing cavity beam $P_o$ by homodyne detection. The van der Waals force acting between the tip apex and a small sample region centered at $x_s$ deforms the probe, which at the tip position is labeled by $w(x,t)|_{x=L}$. Outside of interaction domain ($d \gg a_0$) the probe oscillates at its first resonance frequency $\omega_1$. The PZT actuating the probe base at $z_{s0}$ facilitates the initial probe approach to the surface.}
    \label{setup}
\end{figure}
With reference to Figures \ref{vdw} and \ref{setup}, denoting the distance between the probe tip and the sample surface with $d$, the interaction force $\Gamma = \Gamma (d)$ can be formulated under the assumption of a given contact mechanics for various operational and environmental conditions including charge and thermal states,  pressure, and humidity. We assume that the experimental conditions have been properly adjusted to minimize other contributions (\emph{e.g.}, capillary, electrostatic) to the probe-sample force so that only the van der Waals interaction is relevant. Then, under the conditions of van der Waals-DMT approximation, the force  is modeled~\cite{dmt,garcia} as:
\begin{equation}\label{dmt}
\Gamma(d) = -\alpha \times \begin{cases}
a_0^{-2} - \frac{4}{3}E_f\sqrt{R} (a_0-d)^{3/2} &\text{if $d\le a_0$},\\
d^{-2} &\text{if $d > a_0$},
\end{cases}
\end{equation}
 with $\alpha = HR/6$, where  $H$ is the Hamaker constant ($\sim10^{-20}$~J), $R$ is the tip radius ($\sim 10$~nm), and $a_0$ denotes the distance of the onset of the repulsive part of the interaction ($\approx$ interatomic distance~$\sim 0.165$~nm), and the $E_f$ is the effective probe tip-sample stiffness $E_f^{-1} = (1-\nu_t^2)E_t^{-1} + (1-\nu_s^2)E_s^{-1},$
where the subscripts $t$ and $s$ denote the tip and sample, respectively. 
Example values of the Poisson ratio $\nu$ and Young modulus $E$ are 0.4 and 150GPa, respectively.
  Thus, when  $ d > a_0 $, that is, when the the separation is greater than the interatomic equilibrium distance, $\Gamma$ is dominated by the long range attractive force. Our intent here is not to enter the full contact regime, although the repulsive  force domain can be treated similarly following the results here.  
The force $\Gamma (d)$ as given in Eq.~\ref{dmt} describes the probe-sample interaction as a function of the separation distance  $d$ between the population of the atoms making up the apex of the probe and a number of atoms making up the sample neighborhood immediately below the apex.  However, the typical AFM experiment (\emph{e.g.}, force curve measurement), generates a force as a function of the probe base $z_{s0}$, as shown in Fig.~\ref{setup}.
An approach to the sample surface point $x_s$ of interest reduces $z_{s0}$ and when the probe tip-surface interaction grows beyond a threshold, the cantilever responds by altering its dynamics, balancing the elastic forces with $\Gamma$, with the former depending on probe deformation $w$ and the latter on $d$.    
Noting the transformation from the probe frame ($oxyz$)  to the sample frame:
$\bar{r}_s= \bar{r}_{o}+\bar{r}, \, t_s=t,$ 
and defining the probe base  position
$\bar{r}_{o}=(0,0,z_{s0}(t))$,  for  $d>a_0$ we obtain $d(x_s,t) = z_{s0} + w(x,t)|_{x=L} - h_t,$
where $h_t$ is the tip height, $L$ is the cantilever length, and $w(L,t)$ can be positive or negative. 
The tip height  $h_t$ ($\sim 10~\mu$m), essentially a constant, is for our purpose of no consequence and can therefore be absorbed into $z_{s0}$ by defining $z_0 = z_{s0} - h_t$. Terminating the approach to surface (fixing  $z_{s0}$), $d$  will only fluctuate with $w(L,t)$. We can therefore employ the non-contact interaction force:
\begin{equation}\label{eqg4}
\Gamma_0(t)  = -\frac{\alpha}{d(t)^2}   =  -\frac{\alpha}{[z_0  + w(L,t)]^2}~.
\end{equation} 
Changes in surface properties entering $\Gamma$ through its distance and material dependence can affect  the probe state, both amplitude and phase (\emph{e.g.}, the Hamaker constant can be quantitatively determined~\cite{das}).
We will consider a setup in which the probe is translated from a distance $z_\infty \gg z_0$ to $z_0$ over a finite time. Thus, instead of \eqref{eqg4}, we will use
\begin{equation}\label{eq4ga}
\Gamma(t)  =  \frac{\Gamma_0 (t)}{1+e^{-t/t_0}}~,
\end{equation} 
where $t_0$ is the time over which the translation occurs (around $t=0$). 
As far as the dynamical considerations are concerned, here one may propose alternative interaction forms $\Gamma$, for example, the semi-empirical force employed in the study of nanomechanical frequency mixing~\cite{prl_msafm}.

For frequency of oscillation $\omega_1 \sim 1$~MHz and quality factor $Q \sim 10^4$, the relaxation time of the system is $\tau = Q/\omega_1 \sim 10$~ms. Damping effects can be safely ignored if $t_0 \lesssim \tau$. Therefore, the translation speed (between $z_\infty \sim 100~$nm and $z_0 \ll z_\infty$) ought to be $v \gtrsim 10~\mu$m/s, which is experimentally feasible. Alternatively, one may choose to translate the sample to the probe, an often invoked  maneuver in AFM. 
Initial distance $d$ can be made small enough, so that the transition time during a state squeezing is also small.
The AFM z-piezo can be set to control $d$ (extend time and velocity). The travel distance of the z-piezo can typically be set in the wide nm-$\mu$m range and the  velocity has a range of $\sim 10^2 - 10^5$~nm/s in ordinary systems, and can be $\sim 10^7$~nm/s in specialized systems~\cite{pi}.
Furthermore, the feedback can be engaged or disengaged so that experiments may proceed with or without force feedback control allowing for a variety of measurement flexibilities. When force feedback control is engaged, the tip motion occurs following a preselected force. Since any initial experiments intended to implement QAFM will consider single point measurements rather than raster scanning the sample surface, one may deactivate the feedback. Typically, in force spectroscopy (unlike imaging) the cantilever is the force sensor and the servo feedback loop is deactivated. However, if needed one may activate the feedback loop and reach a satisfactory measurement set-point using ultrafast controllers featuring minimum latency. Current controller technology provides fast feedback (\emph{e.g.}, digital feedback systems~\cite{feed}), while ordinarily, it is possible to use $\sim$~MHz rates fast force-distance measurements (for high extend and retract velocities).

\section{The dynamics of QAFM}
\label{sec:3}
Having defined the form of the tip-sample interaction force,   the probe dynamics may be described by $
\mathcal{C} w(x,t) = \Gamma(t) +F_D(t),$
where $F_D(t)=a_D \cos(\omega_D t +\phi)$ is a harmonic driving force, and $\mathcal{C}$ denotes the operator of the probe dynamics, which can be imported from the Euler-Bernuoulli theory~\cite{ali-george}. We note that here,  in the absence of an applied force, that is, $a_D=0$, in our preliminary study, the oscillating tip-sample distance as given by $d(t)$, except for containing a dc component ($z_{s0}$), is a Brownian process.  
Our aim is to show that $\Gamma$ yields a squeezed quantum state for the probe displacement. We propose to read out such a squeezed state by incorporating the probe as one of the mirrors of a cavity. Typically the top surface of the AFM cantilever probe is vacuum evaporated with a reflective metal thin film ($\approx 50$~nm) for position sensing via laser reflectometry. Minute lateral displacements of the 
sample will induce a variation in $\Gamma$ and a feedback loop subsequently adjusts $d$ to obtain a displacement set-point. Through this process, for example, topographic maps with extremely high resolution are envisioned to be obtained.  

To calculate the normal modes of  the AFM probe, we consider the deformation energy of the microcantilever beam and write  the Lagrangian density: $\mathcal{L} = \mu \dot w^2/2 - u (w, w', w''),$
where $\mu$ is the linear density of the material, and $\dot w = \partial w/\partial t$, $w' = \partial w/\partial x$.
Therefore, the potential energy of the probe is:
\be\label{eq8} U = \int_0^L dx u = \frac{EI}{2} \int_0^L dx \left( w''(x,t)\right)^2   - \frac{\alpha}{z_{0} + w(L,t)}~,\ee
where $I$ is the second moment of inertia. 
The probe equation of motion can be written as:
$ \mu \ddot w + F_1' - F_2'' = F_{\text{int}}, $
where $ F_1 = -\delta U/\delta w',$ $F_2 = -\delta U/\delta w''$, and  $F_{\text{int}} =- \delta U/\delta w$,
subject to  the fixed-free boundary conditions  $w(0,t)=w'(0,t)= w''(L,t) = w'''(L,t) = 0.$
Therefore, the equation of motion is obtained as:
\be\label{eq9} \mu \ddot w +  EI w'''' =   -\frac{\alpha}{[z_{0} + w(x,t) ]^2}\delta (x-L). \ee
\begin{figure}
    \centering
           \includegraphics[width=8.7cm]{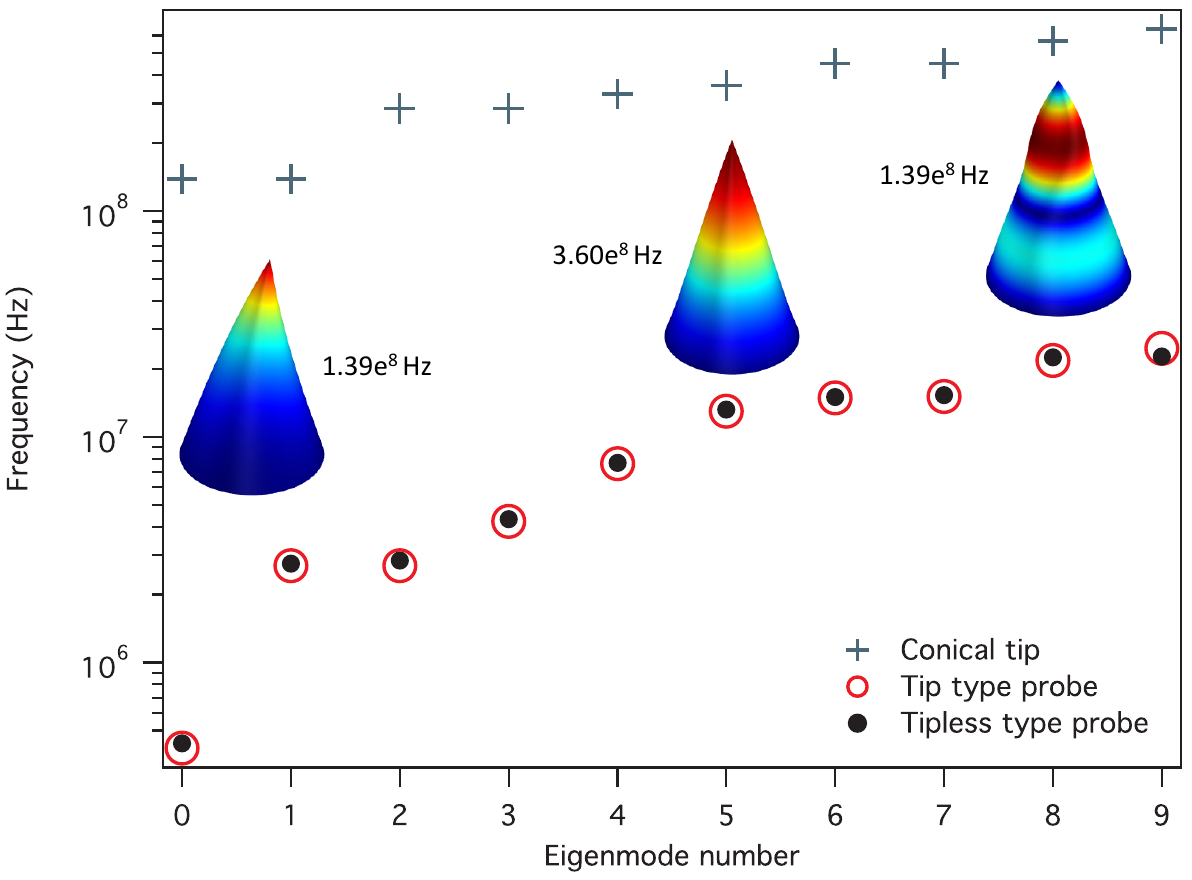}
           \caption[]{   Leading eigenfrequencies of a probe and a probe tip domain as observed from a coordinate system moving with the probe.  
           Three deformation eigenstates of a  conical Si$_3$N$_4$ tip (apex radius of curvature $\sim$ nm, 12~$\mu$m height, 4~$\mu$m radius, and 18$^\circ$  half-angle) are visualized for comparison. Excitation of higher frequency oscillations (higher eigenmode numbers) beyond the first few are less probable. The spectrum of the tip is found to be $\sim$~2 orders of magnitude higher than those of the probe types with the tip and without the tip (100~$\mu$m length, 20~$\mu$m width, 3~$\mu$m thickness). 
                      }
    \label{freqs_modes}
\end{figure}
We note that Eq.\ \eqref{eq9} describes the transversal dynamics of a material domain with uniform mass density per unit length corresponding to a tipless probe, that is, a domain without the subdomain labeled $h_t$ in Fig.~\ref{setup}. This is justified since the dynamics of the tip domain (with a height  $h_t$, Fig.~\ref{setup}) is activated at frequencies at least two orders of magnitude higher than those of the probe (with a length $L$,  Fig.~\ref{setup}). Furthermore, the presence of the tip imposes only  negligible shifts in the frequencies of a tipless type probe. For comparison, we can computationally solve for the  eigenfrequency spectra of a probe with a tip, a probe without a tip, and the tip domain without a probe. We here obtained solutions for a silicon nitride probe with material properties of density $3.1\times 10^3$~kg/m$^3$, Young's modulus $2.5\times 10^{11}$~Pa, and Poisson's ratio 0.23, as shown in Fig.\ref{freqs_modes}. From the computed deformation eigenmodes (see the insets  in Fig.\ref{freqs_modes}), we conclude that while the frequencies of the longitudinal displacements are significantly higher than those of the transversal for the tip domain, both deformation modes are found to exhibit  much larger frequencies than those found for the tip probe beam and the tip-less probe beam. Thus, the dynamics of the tip type probe is primarily dictated by the cantilever and for our purpose the tip can be assumed to amount to an extra mass (as verified computationally by the tip induced frequency shifts in Fig.\ref{freqs_modes}).
With the probe retracted ($d\to \infty$), Eq.~\eqref{eq9} becomes homogeneous, which can be solved in close analogy to  previously reported method~\cite{ali-george} to obtain   
the normalized eigenfunctions $\mathcal{X}_n$, explicitly:
$$ \sqrt{L }\mathcal{X}_n(x) =  \cosh x_n - \cos x_n  - \gamma_n \left( \sinh x_n  - \sin x_n \right), $$
where $x_n = \lambda_n x/L$, and $\gamma_n = \tanh (\lambda_n/2)$, while  $\lambda_n \propto \omega_n$ is a discrete set of eigenvalues. 
Though not of interest in this work, we note that for each $\omega_n$, the sensitivity of probe's frequency shift varies with sample surface stiffness~\cite{turner,chang}.
For the inhomogeneous  case, considering the nonlinearity in Eq.~\ref{eq9}, though a general solution  may be obtained seminumerically from an eigenfunction expansion~\cite{prl_msafm}, we may formally write  
\begin{equation}
w(x,t)= \sum_{nk} Q_{nk}(t) \mathcal{X}_n(x)  e^{(\rho_{nk}-i\omega_{nk}) t},
\label{fullsol}
\end{equation}
where the amplitude polynomials $Q_{nk}(t)$ must be  determined for each spatial eigenmode $n$, for which there is an infinite number of eigenvalues $(\rho_{nk},\omega_{nk})$ due to the nonlinearity~\cite{msafm,prl_msafm}. 
Far away from the surface, 
the probe may be assumed to be in the lowest mode, where
the state of the cantilever beam is approximately $ w(x,t) \approx \sqrt{L} q(t) \mathcal{X}_1(x).$
When approaching the surface (sufficiently small $d$),  the attractive force  introduces a non-linear correction. 
In this limit, denoting  the mass of the cantilever with $m= \mu L$ (numerically, $m\approx 30$~ng),
we integrate the potential and obtain:
\be U = \frac{1}{2} m \omega_1^2 q^2  +  \frac{\alpha \sqrt{L} \mathcal{X}_1 (L)}{z_{0}^2} q - \frac{\alpha L \mathcal{X}_1^2 (L)}{z_{0}^3} q^2 + \mathcal{O}(q^3), \ee
from which, we obtain the minimum of the potential   $U'(\bar{q}) = 0$
so that
small oscillations around the stable equilibrium point $q=\bar{q}$  have frequency $\Omega$, given by:
\be\label{eqOmega} \Omega^2 = \omega_1^2 - \delta^2,\ \  \ \ \ \ \delta^2 = \frac{2\alpha L \mathcal{X}_1^2(L)}{mz_{0}^3} = \frac{8\alpha }{mz_{0}^3}~,
 \ee
 where the last term is obtained by noting $\mathcal{X}_n (L) =  2L^{-1/2}$.
For typical values of relevance to AFM~\cite{biggs,garcia,prl_msafm}, Fig.~\ref{omega} displays the region where $\Omega$ is severely suppressed. Thus, for a probe oscillating freely (\emph{e.g.}, following an excitation to its dominant eigenmode), its frequency is reduced when it enters the van der Waals attractive force regime. We note that ordinarily, while significant spectral shifts to lower frequencies are observed in experiments when approaching the surface, the limit of no oscillation ($\Omega = 0$) is not reached for all eigenmodes due to the presence of noise that drives the system and invokes the first few eigenmodes. Oscillations at the reduced frequencies $\Omega$ occurs with an amplitude proportional to: 
 \begin{equation}\label{qbar}
\bar{q} = \frac{2\alpha z_0}{8\alpha -m\omega_1^2 z_0^3}~.
\end{equation}
Considering that the transversal deformations provide the primary oscillator mode in AFM, assuming a proper calibration $\delta w \sim \delta q \sim \delta S$, where $S$ is the signal, one may attempt a sensitivity analysis from the phase and amplitude of $w$. The QAFM sensitivity may be expressed as the minimum $\delta q$ that yields a signal $\delta S$ from the cavity readout that is of the same order of magnitude as that of the noise.  A change $\delta q$ can occur as a result of a sample translation ($\delta x_s$ in Fig.~\ref{setup}), a change in sample material response ($\delta \alpha$ entering $\Gamma$), or a change in vertical translation ($\delta z_0$ entering $\Gamma$).

   
\section{Quantum squeezed atomic force microscopy}\label{sec:q}

The quantum description of the proposed metrology is here based on the assumption that at low temperatures, the probe, initially outside of the interaction regime,  behaves approximately as a quantum harmonic oscillator of frequency $\omega_1$ with the Hamiltonian:
\be\label{eqHb1} H = \frac{p^2}{2m} + \frac{1}{2} m\omega_1^2 q^2 = \hbar\omega_1 \left( b_1^\dagger b_1 + \frac{1}{2} \right), \ee
where $b_1$  annihilates  the ground state $|0\rangle_{\omega_1}.$ The probe operators $q, p$ satisfy $[q,p] = i\hbar$ with negligible quantum fluctuations of the displacement $\Delta q \sim   (\hbar/m\omega_1)^{1/2}  \sim 0.1$~fm.
Approaching the surface, the probe is translated to a position $z_0$, such that  the stable equilibrium point  
is shifted to $q=\bar{q},$ near which small oscillations occur at frequency $\Omega$,  given by Eq.\ \eqref{eqOmega}. 
With the probe translated to within the attractive regime of the probe-sample interaction, the Hamiltonian in the Hilbert space of no excitations of higher flexural modes becomes
\be\label{eqHa} H = \frac{p^2}{2m} + \frac{1}{2} m\Omega^2 (q - \bar{q})^2 = \hbar\Omega \left( a^\dagger a + \frac{1}{2} \right)~, \ee
where the annihilation operator $a$, satisfying $[a, a^\dagger] =1$,   is now defined by
\be\label{eq32} a = -  \zeta  +  \frac{1}{\sqrt{2\hbar m \Omega}} \left[ m \Omega q + ip \right], \ \
 \zeta = \sqrt{\frac{m \Omega}{2\hbar}} \bar{q} , \ee
instead of $b_1$  in Eq.\ \eqref{eqHb1}. 
With the probe within the interaction region,  its ground state $|0\rangle_\Omega$ is annihilated as $a|0\rangle_\Omega =0$,
while when retracted outside of the region, the Hamiltonian \eqref{eqHa} 
reduces to \eqref{eqHb1}.  
Entering the deeper part of the attractive regime of the probe-sample interaction, the quantum fluctuations of the probe displacement  $\Delta q\sim 0.1 \sqrt{\omega_1/\Omega}$~fm diverge when $\Omega \to 0$. 
From the definitions of the probe operators $(p,q)$ and Eq.~\eqref{eq32}, we observe that the operators $a$ and $b_1$ annihilating the ground state of the interacting probe  $|0\rangle_{\Omega},$ and the free probe $|0\rangle_{\omega_1},$ respectively, are related. The modes in the two cases, that is, the probe near the sample surface and away from it, are related through the Bogoliubov transformation:
\be b_1 = \cosh r (a+\zeta  ) + \sinh r (a^\dagger + \zeta ), \ \ \ \ e^{2r} = \frac{\omega_1}{\Omega} > 1,  \ee
where $r$ is the squeezing parameter.
From this result, Eqs.~\eqref{qbar}, and \eqref{fullsol}, assuming  $\alpha$ does not change when $d$ is varied, and denoting $\tilde{r} =e^{4r}$, one may implicate the role of the squeezing in a measure of  sensitivity of the QAFM from the following observations:
\begin{equation}
\Eval{\frac{\partial \bar{q}}{\partial z_0}}{\alpha}{} = \frac{1}{4}\left (  1- 4 \tilde{r} + 3\tilde{r} ^2  \right ), 
 \Eval{ \frac{\partial \bar{q}}{\partial \alpha}}{z_0}{} =  \left[ \frac{ \tilde{r} \left (\tilde{r}  -1 \right )^2} {216 m \alpha^2\omega^2} \right ]^{1/3},   
\label{sens}
\end{equation} 
\begin{figure}
    \centering
                 \includegraphics[width=8.0cm]{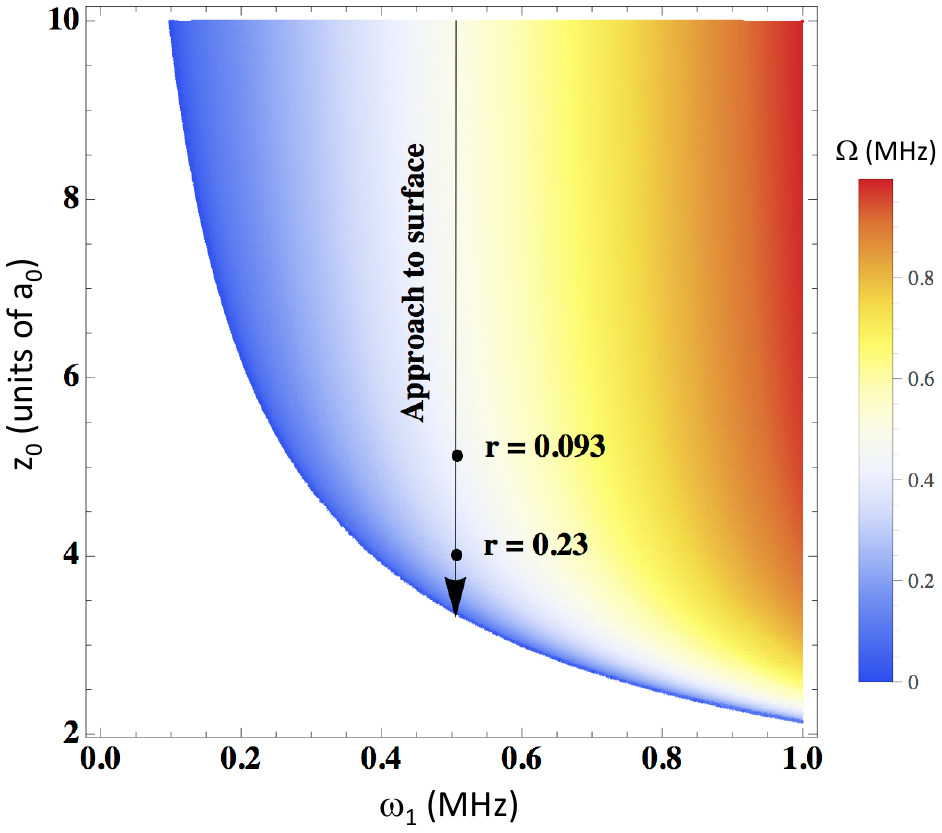}
           \caption[]{Domain of small oscillation frequency $\Omega$ around the stable equilibrium point $q=\bar{q}$. For a given unengaged ($d\to \infty$) probe frequency,  say the first  resonance $\omega_1$, the oscillation frequency is severely reduced  when the probe approaches the surface ($z_0 \approx w$). For a probe with a free frequency of 0.5~MHz, an approach to the surface leads to squeezing as exemplified by the two values of the squeezing parameter $r$.}
    \label{omega}
\end{figure}
both of which increase with squeezing.
The ground state of the probe, within the van der Waals regime, $|0\rangle_\Omega$, is a \emph{squeezed coherent} state in the system with the probe away from the sample surface~\cite{bibTapia}.  When the probe operates near the sample surface, $r$ is large.
Therefore, $r$ can change dramatically, as the probe retracts from the sample surface. This is evident from the map showing the probe frequency shift in Fig.~\ref{omega}, where the arrow depicts the approach to the surface. As can be seen, for a given unperturbed probe frequency $\omega_1 = 0.5$~MHz, the squeezing effect associated with the nonlinear force, as exemplified by the two values of the squeezing parameter $r$ for two probe-sample distances, can be significant. We will now proceed to obtain the uncertainties associated with the probe displacement quadratures.        
In QAFM (Fig.~\ref{setup}), the output field of the cavity is measured by homodyne detection from the current output of the photodiode detectors converted to an output voltage. The detected  signal $S$ will be proportional to the fluctuations of the quadrature component of the output field in phase with the local oscillator. From a calibration of $S$, a conversion to displacement can be achieved.

Achieving the squeezed state in QAFM, while ideal when the probe is in its ground state, does not require it. It has been shown~\cite{chan} that a micromechanical oscillator can be prepared in  its ground state, although quite often it is prepared in an initial state which is close to the ground state. Following  cryogenic and optomechanical cooling, this state has been demonstrated~\cite{metzger,Kleckner,Karrai,park,oconnell,teufel,grob}.
Suppose that initially the probe is retracted, that is, far away from the sample surface, and the beam has been cooled down to a temperature $T$ in the mK range following the quantum state preparation protocol of pulsed optomechanics~\cite{bibpulse,vanner,ali-george}. 
The initial state prepared via application of short laser pulses is detected using homodyne detection in the cavity as depicted in Fig.~\ref{setup}.
Here, we note that other experimentally demonstrated methods including 
a combination of precooling, feedback cooling, sideband cooling and cryogenic cooling~\cite{bibpulse,park}
 may be suitable as well. Also noteworthy is that the AFM is compatible with such operational conditions and low temperature AFM applications involving operation at  $T\sim 1$~K, without invoking cavity cooling, have been reported for a variety of  investigations~\cite{Hembacher,Prior,Bhatt,Thomson,Bhatt2,Arima}.

Having prepared  the system  in the thermal state with an average number of phonons
$\langle n \rangle \approx k_BT/\hbar\omega_1$ (for $\omega_1/2\pi = 1$~MHz, $T = 10$~mK, we have $\langle n\rangle \approx 220$),
we will now reduce $d$ either by the probe approaching the sample surface or by the sample approaching the probe (both modalities available in AFM technology) and study the response of the probe. 
Thus, defining $c_1 = (\hbar/2m\omega_1)^{-1/2}$ and $c_2 = -i(2/\hbar m\omega_1)^{1/2}$
we obtain, for the free probe, a state with quadratures and their respective uncertainties:
\bes\label{eq25}  X^0_1 &=&   b_1^\dagger + b_1 = c_1q, \,\, \Delta  X^0_1 = \sqrt{ \frac{\chi^2 +\beta_+ }{2(1 + \beta_+ \chi^2 + \chi^4)}},\nonumber\\
X^0_2 &=& b_1^\dagger - b_1 =c_2 p, \,\,   \Delta X^0_2= \sqrt{ \frac{1 +\beta_- \chi^2 + \chi^4}{2(\chi^2 + \beta_-)}},  \ees
where $\chi$ is a coupling constant, and 
$\beta_{\pm} =   (1 \pm \beta)/(1 \mp \beta),$ with $\beta = \exp{(-\hbar\omega_1/k_BT)}.$
Similar to the case of cavity coupling to an electrostatically actuated microbeam~\cite{ali-george}, when the probe is decoupling from the cavity ($\chi  \to 0$), the uncertainties match those of the initial thermal state of the probe: 
$\rho = (1-\beta) \sum_{n=0}^\infty \beta^n |n\rangle\langle n|,$
where the states $|n\rangle$ are created with $b_1^\dagger$. 
On the other hand, for strong coupling ($\chi\to\infty$), or even at moderate coupling ($\chi \sim 1$), the state approaches a minimum uncertainty wave packet~\cite{ali-george}.

\begin{figure}
    \centering
           \includegraphics[width=8.5cm]{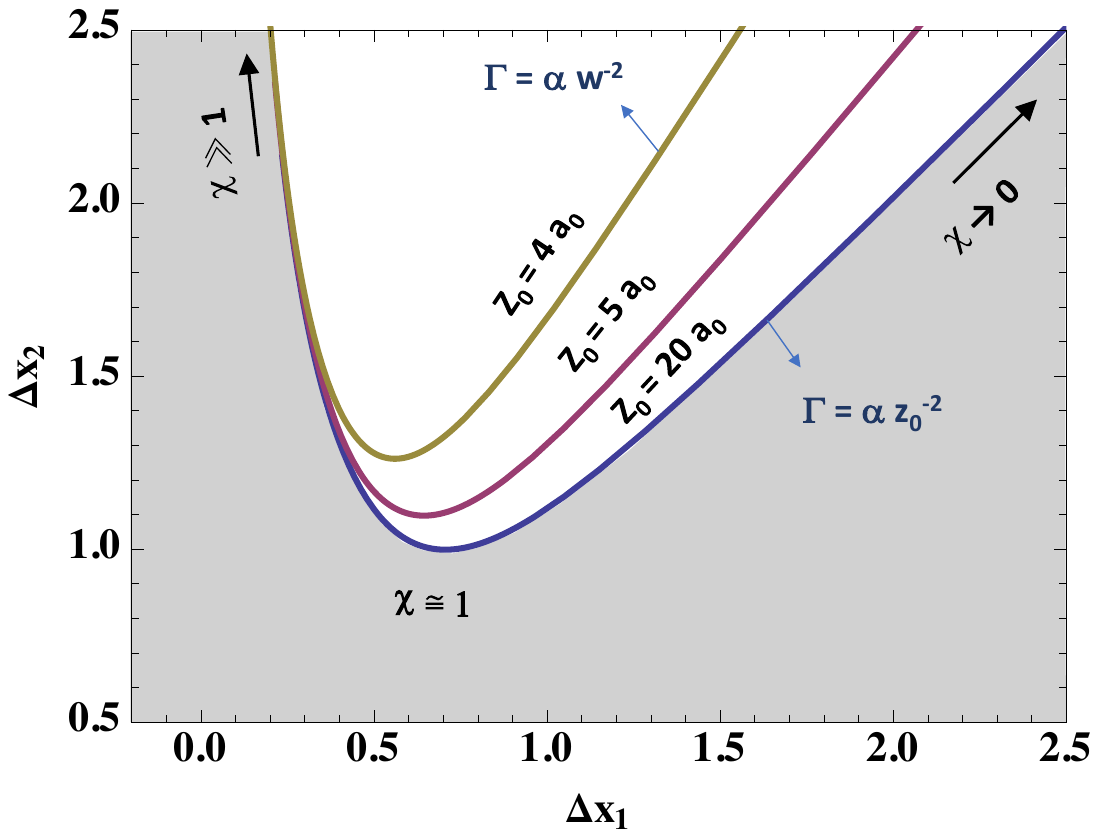}
           \caption[]{The parametric behavior of the probe uncertainties. The gray shaded region represents the free ($z_0\gg a_0$) cantilever domain ($\Gamma \approx \alpha z_0^{-2} \to 0$). For sufficiently small probe-sample distances, the uncertainties in the quadratures is dictated by the strength of the coupling to the cavity, starting from the top right of the blue curve for a weakly coupled probe and tracing the top left segment of the blue curve for a strong coupling. For smaller probe-sample separation, the nonlinear force induced squeezing yields a tighter distribution of the uncertainties, as shown by the inner trace for $z_0 = 4 a_0$, after which the force takes the asymptotic form of $\Gamma \approx \alpha w^{-2}$. A probe oscillation frequency of 0.5~MHz was assumed (bandwidth $\sim 10^2 - 10^3$~Hz).} 
    \label{param}
\end{figure}

With the probe prepared in a state close to the ground state, we suddenly ($t_0 \sim \tau $) reduce $d$ by translating  the probe (sample surface) close to the sample surface (probe).
The initial state of the probe is built on the ground state $|0\rangle_{\omega_1}$, which is not annihilated by $a$ (see Eq.\ \eqref{eq32}). Instead, it is a \emph{squeezed} state.
After the transition, the relevant quadratures and their respective uncertainties are given by:
\bes\label{eq61} X_1 &=&  \frac{1}{\sqrt{2}} \left( a^\dagger + a \right) = e^{-r} X^0_1 \,\,\, \Rightarrow \,\,\,   \Delta X_1 = e^{-r} \Delta X^0_1  \ , \nonumber\\
X_2 &=& \frac{i}{\sqrt{2}} \left( a^\dagger - a \right) = e^{r} X^2_2  \,\,\, \Rightarrow \,\,\,  \Delta X_2 = e^{r} \Delta X^0_2.  \ees
As the probe approaches the sample surface, we have $\Omega \to 0$, therefore, $\Delta X_1$ can be made very small and $\Delta X_2$ very large, while the product $\Delta X_1 \Delta X_2$ remains near minimum uncertainty. The behavior of the uncertainties is studied in Fig.~\ref{param} in the cavity coupling strength and probe-sample separation distance parameter domain. 
To study this effect in detail, we can model the time-dependence of the external force by Eq.\ \eqref{eq4ga}.
This can be carried out in a precise manner as for the case  of 
squeezing near the pull-in instability of an electrostatically biased  beam~\cite{ali-george}. Thus,  regarding the Heisenberg equations of motion, we can revisit the development following Eq.~(67) in the recently reported calculation~\cite{ali-george}.

To estimate  the lifetime
of the squeezed state $\tau_s$ due to its interaction (dissipation/thermalization) with the environment, we first note that the probe frequency and the decay time of the cavity satisfy $\omega_1 \tau_c \ll 1$, and thus the position of the probe  will not change appreciably during $\tau_c$. While the majority of experiments within cavity optomechanics employ continuous monitoring of the mechanical position, the pulsed protocol assumed in the description of the QAFM, owing to its pulsed nature and short time scales, is not as susceptible to the thermal bath, as it can be performed on short time scales~\cite{vanner_natcom}. 
Considering the long relaxation time of the oscillator in response to any transient excitation, an estimate may be obtained for the squeezed state based on the evolution of the $q(t)$. Furthermore, from the imaginary part of a solution for $q$ obtained in a similar fashion as for the case of an electrostatically biased oscillator~\cite{ali-george}(see Eq.~70 therein), we may estimate a decay time for the probe under the influence of $\Gamma$. The worst-case scenario of having a prepared probe in the squeezed state experiencing a sudden loss of squeezing mechanism, we expect a life time $\tau_s \sim \tau \sim$~ms. This is consistent with the estimate for the effect of the bath on the prepared state via pulsed optomechanics (see Eq.~8 in Vanner et al.~\cite{bibpulse}). Using this estimate, for the parameters assumed in our work, one observes an initially squeezed variance will grow to 0.5 on times $\sim$~ms.       
The longer relaxation time of the oscillator can ensure that measurements of the squeezing effects can be carried out without instabilities associated with oscillation distortion or damping effects. The higher $Q$ may also be preferred in case one chooses to invoke a feedback loop. Modern ultrafast piezo transducers and feedback systems can deliver better control when the response time of the probe is shorter.
Probes with stiffness $> 40$~N/m or $< 0.01$~N/m are typically employed in AFM, yielding a broad range of $Q$, which also affects the sensitivity (\emph{e.g.,}  a probe with a stiffness of 28~N/m and $\omega_1 = 318$~kHz, exhibits an amplitude spectral density of 342~$\text{fm}/\sqrt{\text{Hz}}$ for the thermal noise~\cite{garcia}).

To estimate the displacement sensitivity for a given probe, we first assume a very high finesse cavity and note $\delta S \propto \delta q$, a wavelength of  $\lambda \sim \mu$m, and  a probe with $\omega_1 = 0.5$~MHz, to obtain (following Hadjar et al.~\cite{hadjar}) 
$\sim 10^{-19}$~m/$\sqrt{\text{Hz}}$. 
For lower finesse cavities~\cite{vanner_natcom} (\emph{e.g.}, $\sim 10^3-10^4$), sensitivities $\lesssim 10^{-18}$~m/$\sqrt{\text{Hz}}$ are expected. Furthermore, with sensitivities $\sim \text{fm}/\sqrt{\text{Hz}}$ already demonstrated by pulsed optomechanics~\cite{vanner_natcom}, in light of the described van der Waals force induced squeezing and Eq.~\eqref{sens}, we  expect the QAFM to improve the measurement sensitivity by several orders of magnitude above $\sim \text{fm}/\sqrt{\text{Hz}}$.

Finally, we note that the cavity is not essential to achieve squeezing in QAFM but to measure it, which differentiate the presented approach from  methods such as ponderomotive squeezing~\cite{fabre}  or mechanical squeezing via parametric resonance~\cite{param}, where either the optical field is squeezed or a modulated input field is essential for the squeezing.
However, a comparison of the presented method with that of Fabre et al.~\cite{fabre} who explored the optical interaction in a Kerr medium and the further noise reduction in the optical field near the bistability turning points may be interesting. 
The mechanical squeezing in our case near the stability region near the pull-in parameter domain considered in our previously reported electrostatically biased microbeam~\cite{ali-george} as well as in the case of QAFM could potentially be invoked in a ponderomotive squeezing. Similarly, the concept of squeezing via periodically modulated driving field and the parametric resonance by Liao and Law~\cite{param}  is of relevance since the underlying dynamics is directly accessible within the response of cantilevers. In particular, the assumption of a single cavity mode field, the modeling of the moving mirror as a harmonic oscillator, and the inclusion of the damping by coupling the system with an oscillator bath (yielding quantum Langevin equation) are reminiscent of similar assumptions made in our current and previous work~\cite{ali-george}. Further fruitful analogies between these oscillators may be noted since once they are prepared in an initial state, one experiences an exponential squeezing in time and the other (our case) undergoes a squeezing with distance (implicitly with time) but of course due to two different kinds of forces (van der Waals in our case). However, in both cases of ponderomotive and parametric resonance the cavity input field is essential in the observed squeezing: the cavity output field in one case~\cite{fabre} and the mechanical state in the latter~\cite{param}. 

\section{Conclusions and outlook}\label{sec:conc}
In summary, we presented a new measurement modality suitable for implementation in scanning probe microscopy
and in sensing where one object interacts nanomechanically with a second object. 
The quantum state atomic force microscopy, constituting a new concept in quantum metrology, was here formulated for the first time and explored within the van der Waals force regime. However, other forces of relevance, such as electrostatic, magnetostatic, Casimir, etc., possessing different nonlinearities, may be modeled in a similar fashion following the presented results.   
Here, the proposed nanometrology capitalizes on the nanomechanical interaction between a probe and a sample surface in the quantum regime of the probe deformation state. 
While, for the sake of simplicity and without loss of generality, the feasibility of the quantum enabled  force microscopy was studied only in the attractive regime of the probe-sample interaction and for the fundamental probe frequency, a similar treatment can be obtained for the full force regime and higher mechanical eigenfrequencies  following the presented results. While the QAFM concept certainly warrants further investigation, within this introductory effort the vdw-DMT was shown to enable the necessary amplitude squeezing for the QAFM to be a viable approach to scanning probe microscopy with potential to deliver a resolving power significantly beyond pm.     
The cavity optomechanical readout of QAFM in conjunction with the capitalization of the interfacial forces to squeeze the mechanical motion of the probe is of potential to lead to many interesting applications such as parametric cooling, while being metrologically compatible with other harnessing mechanisms such as feedback cooling.    
In addition to  providing a dynamic platform for exploring the effect of various regimes of the interfacial forces, the key benefits of the QAFM include applications within force spectroscopy with significantly improved resolution in measurements of electronic, spin, and magnetic properties of molecular and atomic systems, clusters, and quantum dots and their sub-component properties.  Other force sensing investigations motivated by a need for ultra-sensitive single-molecule force measurements are underway (\emph{e.g.}, see Doolin \emph{et al}~\cite{doolin}).

\begin{acknowledgments}
This research was supported in part by the laboratory directed research and development fund and in part 
by the BioEnergy Science Center (BESC) of the Oak Ridge National Laboratory (ORNL). The BESC is a US Department of Energy (DOE) Bioenergy Research Center supported by the Office of Biological and Environmental Research in the DOE Office of Science.
ORNL is managed by UT- Battelle, LLC, for the US DOE under contract DE-AC05- 00OR22725.
\end{acknowledgments}


\begin{thebibliography}{}

\bibitem{pico}
 R. Sharma and  A. Sharma, C. J. Chen, 
 \emph{Emerging Trends of Nanotechnology towards Picotechnology: Energy and Biomolecules,}
 Nature Precedings, doi:10.1038/npre.2010.4525.1 (2011).
%
\bibitem{msafm}
 L.\ Tetard, A.\ Passian, and T.\ Thundat, \emph{New modes for subsurface atomic force microscopy through nanomechanical coupling,} Nature Nanotech. {\bf 5},
105 (2009).
%
\bibitem{hpfm}
 L.\ Tetard, A.\ Passian, R.\ H.\ Farahi, T.\ Thundat, and B.\ H.\ Davison, \emph{Opto-nanomechanical spectroscopic material characterization,} Nature Nanotechnology {\bf 10},
870 (2009).
%
\bibitem{rugar}
H.\ J.\ Mamin and D.\ Rugar, \emph{Sub-attonewton force detection at millikelvin temperatures,} Appl.\ Phys.\ Lett.\ {\bf 79}, 3358 (2001).    
%
\bibitem{moser}
J.\ Moser, J.\ G\"uttinger, A.\ Eichler, M.\ J.\ Esplandiu, D.\ E.\ Liu, M.\ I.\ Dykman, and A.\ Bachtold, \emph{Ultrasensitive force detection with a nanotube mechanical resonator,} Nature Nanotechnology  {\bf 8}, 493 (2013).
%
\bibitem{gavartin}
E.\ Gavartin, P.\ Verlot, and T.\ J.\ Kippenberg, \emph{A hybrid on-chip optomechanical transducer for ultrasensitive force measurements,} Nature Nanotechnology {\bf 7}, 509 (2012).
%
\bibitem{ranjit}
G.\ Ranjit, M.\ Cunningham, K.\ Casey, and A.\ A.\ Geraci, \emph{Zeptonewton force sensing with nanospheres in an optical lattice,} Phys.\ Rev.\ A {\bf 93}, 053801 (2016).
%
\bibitem{aspelmeyer}
M.\ Aspelmeyer, T.\ J.\ Kippenberg, and F.\ Marquardt, \emph{Cavity optomechanics,} Rev.\ Mod.\ Phys.\ {\bf 86}, 1391 (2014).

\bibitem{walls:milburn}
D.\ F.\ Walls and  G.\ J.\ Milburn,  \emph{Quantum Optics}, Springer, Berlin (2008).

\bibitem{nat}
D. W. C. Brooks, T. Botter, S. Schreppler, T. P. Purdy, N. Brahms, and D. M. Stamper-Kurn, \emph{Non-classical light generated by quantum-noise-driven cavity optomechanics,} Nature {\bf 488}, 476  (2012).

\bibitem{ligo}
The LIGO Scientific Collaboration, \emph{Enhanced sensitivity of the LIGO gravitational wave detector by using squeezed states of light,}  
Nature Photonics {\bf 7}, 613 (2013).

\bibitem{taylor}
M. A. Taylor, J. Janousek, V. Daria, J. Knittel, B. Hage, H. A. Bachor, and W. P. Bowen, 
\emph{Biological measurement beyond the quantum limit,} Nature Photonics {\bf 7}, 229  (2013).

\bibitem{lecocq}
F. Lecocq, J. B. Clark, R. W. Simmonds, J. Aumentado, and J. D. Teufel,  
\emph{Quantum Nondemolition Measurement of a nonclassical state of a massive object,} Phys. Rev. X {\bf 5}, 041037
(2015).
%
\bibitem{dmt}
B.\ V.\ Derjaguin, V.\ M.\ Muller, and Y.\ P\ Toporov, \emph{Effect of contact deformations on the adhesion of particles}, Journal of Colloid and Interface Science, {\bf 53}, 314 (1975).   

\bibitem{garcia}
R. Garcia, \emph{Amplitude modulation atomic force microscopy}, Wiley-VCH (2010).

\bibitem{das}
S. Das, P. A. Sreeram, and A. K. Raychaudhuri, \emph{A method to quantitatively evaluate the Hamaker constant using the jump-into-contact effect in atomic force microscopy,} Nanotechnology {\bf 18}, 035501 (2007).

\bibitem{prl_msafm}
L.\ Tetard, A.\ Passian, S.\ Eslami, N.\ Jalili, R.\ H.\ Farahi, and T.\ Thundat, \emph{Virtual Resonance and Frequency Difference Generation by van der Waals Interaction,} Phys.\ Rev.\ Lett.\ {\bf 106}, 180801 (2011).

\bibitem{pi}
Physik Instrumente, www.pi-usa.us
\bibitem{feed}
Anfatec Instruments AG, www.anfatec.com
%
\bibitem{ali-george}
A.\ Passian and G.\ Siopsis, \emph{Strong quantum squeezing near the pull-in instability of a nonlinear beam}, Phys.\ Rev.\ A {\bf 94}, 023812 (2016).

\bibitem{turner}
J. A. Turner and J. S. Wiehn, \emph{Sensitivity of flexural and torsional vibration modes of atomic force microscope cantilevers to
surface stiffness variations,}
Nanotechnology {\bf 12}, 322 (2001).
\bibitem{chang}
W. J. Chang,  \emph{Sensitivity of vibration modes of atomic force microscope cantilevers in continuous surface contact,}
Nanotechnology {\bf 13}, 510 (2002).

\bibitem{biggs}
S.\ Biggs and P.\ Mulvaney, \emph{Measurement of the forces between gold surfaces in water by atomic force microscopy,} J.\ Chem.\ Phys.\ {\bf 100}, 8501 (1994).
%
\bibitem{bibTapia}
R.\ Mu\~noz-Tapia, \emph{Quantum mechanical squeezed state,} Am.\ J.\ Phys.\ {\bf 61}, 1005 (1993).

\bibitem{chan}
J. Chan, T. P. Mayer Alegre, A. H. Safavi-Naeini, J. T. Hill, A. Krause, S. Gr\"oblacher, M. Aspelmeyer, and O. Painter, Nature {\bf 478} 89 (2011).
\bibitem{metzger}
C. H\"ohberger Metzger, and K. Karrai, Nature {\bf 432}, 1002 (2004).
\bibitem{Kleckner}
D. Kleckner, and D. Bouwmeester, Nature {\bf 444}, 75 (2006).
\bibitem{Karrai}
I. Favero, and K. Karrai,  Nature Photonics {\bf 3}, 201 (2009).
\bibitem{park}
Y. S. Park and H. Wang, \emph{Resolved-sideband and cryogenic cooling of an optomechanical resonator,}
Nature Physics {\bf 5}, 489 (2009).
%
 \bibitem{oconnell}
 A. D. OÕConnell et al., \emph{Quantum ground state and single-phonon control of a mechanical resonator,}
 Nature {\bf 464}, 697 (2010).
 \bibitem{teufel}
 J. D. Teufel et al.,  \emph{Sideband cooling of micromechanical motion to the quantum ground state,}
Nature {\bf 475}, 359 (2011).
 \bibitem{grob}
S. Gr\"oblacher,  et al., \emph{Demonstration of an ultracold micro-optomechanical oscillator in a cryogenic cavity,} 
Nature Physics {\bf 5}, 485 (2009).

\bibitem{bibpulse}
M.\ R.\ Vanner, I.\ Pikovski, G.\ D.\ Cole, M.\ S.\ Kim, \v{C}.\ Brukner, K.\ Hammerer, G.\ J.\ Milburn, and M.\ Aspelmeyer, \emph{Pulsed quantum optomechanics,}
PNAS {\bf 108}, 16182 (2011).
\bibitem{vanner}
M.\ R.\ Vanner, I.\ Pikovski, and M.\ S.\ Kim, \emph{Towards optomechanical quantum state reconstruction of mechanical motion,} Ann.\ Phys.\ {\bf 527},  15  (2015).
%
\bibitem{Hembacher}
 S. Hembacher, F. J. Giessibl, J. Mannhart, and C. F. Quate, \emph{Revealing the hidden atom in graphite by low-temperature atomic force microscopy,} PNAS {\bf 100} 12539 (2003).

\bibitem{Prior}
M. Prior, A. Makarovski, and G. Finkelstein, \emph{Low-temperature conductive tip atomic force microscope for carbon nanotube probing and manipulation,} Appl. Phys. Lett. {\bf 91}, 053112 (2007).

\bibitem{Bhatt}
P. M. Bhatt, D. H. Agrawal, A. M. Pathan, and N. D. Chauhan, \emph{In2o3 Thinfilm Analysis by Low Temperature Atomic Force Microscope,} Procedia Engineering {\bf 38}, 661 (2012).

\bibitem{Thomson}
R. E. Thomson, \emph{Low-temperature atomic force microscope using piezoresistive cantilevers,} Rev. Sci. Inst. {\bf 70},  3369 (1999).
%
\bibitem{Bhatt2}
P. M. Bhatt, U. S. Joshi, H. N. Shah, P. K. Brahmbhatt, \emph{Low temperature Atomic Force Microscopy-A Review,} IOSR Journal of Mechanical and Civil Engineering (IOSRJMCE) ISSN : 2278-1684 {\bf 2(3)} , 01 (2012).

\bibitem{Arima}
E. Arima, H. Wen, Y. Naitoh, Y. J. Li, and Y. Sugawara, \emph{Development of low temperature atomic force microscopy with an optical beam detection system capable of simultaneously detecting the lateral and vertical forces,} Rev. Sci. Inst. {\bf 87}, 093113 (2016).
%
\bibitem{vanner_natcom}
M. R. Vanner, J. Hofer, G. D. Cole, and M. Aspelmeyer, \emph{Cooling-by-measurement and mechanical state tomography via pulsed optomechanics,} Nature Communications {\bf 4} 2295 (2013).

\bibitem{fabre}
C. Fabre, M. Pinard, S. Bourzeix, A. Heidmann, E. Giacobino, and S. Reynaud
\emph{Quantum-noise reduction using a cavity with a movable mirror,} Phys. Rev. A {\bf 49}, 1337 (1994).
%
\bibitem{param}
J. Q. Liao and C. K. Law, \emph{Parametric generation of quadrature squeezing of mirrors in cavity optomechanics,} 
Phys. Rev. A {\bf 83}, 033820 (2011).

\bibitem{hadjar}
Y. Hadjar, P. F. Cohadon, C. G. Aminoff, M. Pinard, and A. Heidmann, \emph{High-sensitivity optical measurement of mechanical Brownian motion} 
Europhys. Lett. {\bf 47} 545 (1999).

\bibitem{doolin}
C. Doolin, P. H. Kim, B. D. Hauer, A. J. R. MacDonald, and J. P. Davis, \emph{Multidimensional optomechanical cantilevers for high-frequency force sensing,} New Journal of Physics {\bf 16}, 035001 (2014). 

\end{thebibliography}

\end{document}